\documentclass[namedreferences]{kluwer}
\usepackage{graphicx}
\begin{document}
\begin{article}
\begin{opening}
\title{Numerical simulations of self-gravitating magnetized disks}            

\author{S. \surname{Fromang}\email{fromang@iap.fr}}
\institute{Institut d'Astrophysique de Paris}                               
\author{J.P \surname{de Villiers}}
\author{S.A \surname{Balbus}}
\institute{University of Virginia, USA}




\runningtitle{MHD simulations of self-gravitating disks}
\runningauthor{Fromang,S. et al}



\begin{abstract} 
We present the first global simulations of self-gravitating magnetized
tori. The simulations are performed with Zeus-2D and GLOBAL. We
find the magnetorotational instability (MRI) to behave similarly
in a self-gravitating environment as in previous simulations of non
self-gravitating systems: enhancement of turbulent angular momentum
transport follows the linear phase. The torus quickly develops a two
component structure composed of an inner thick disk in Keplerian rotation
and an outer massive disk. We compare this result with zero mass global
simulations in 2D, and also present preliminary results of 3D simulations.
\end{abstract}




\end{opening}

\section{Introduction}

In the early phases of star formation, the forming accretion disks are
likely to be very massive because of a strong infall from the parent
molecular cloud. These massive disks are subject to the development of
gravitational instabilities which redistribute angular momentum
\cite{laughlin98}. However, when sufficiently ionized, the disks are also
unstable to the MRI \cite{balbus91,balbus98}. The simultaneous development
of both instabilities in these disks may significantly affect their 
evolution. We have therefore undertaken a numerical study of this
phenomenon by means of numerical simulations of self-gravitating
magnetized tori.\\
In section 2, we present a summary of our numerical methods, and describe
the initial equilibrium configuration. In section 3, we present the
results of the 2D simulations, and compare them with nonself-gravitating
tori evolution. In section 4, we review preliminary results obtained in
3D, and we discuss the future developments of this work in section 5.

\section{Numerical methods}

\subsection{Algorithms}
We used the code Zeus-2D \cite{Stone92a} to perform the axisymmetric
calculations in cylindrical coordinates. Zeus-2D solves the MHD equations
using time-explicit Eulerian finite differencing. Magnetic fields are
updated with the Constrained Transport method \cite{Evans88} in order to
preserve $\nabla \cdot \vec B = 0$ and the method of characteristics is
used to compute the electromotive forces in order to accurately describe
the propagation of Alfven waves. The Poisson solver in Zeus-2D has been
modified and involves two steps.  We first calculate the gravitational
potential $\phi_g$ on the boundary, using the Legendre functions
well-suited to our cylindrical geometry \cite{Cohl99}, and we then apply the
Successive Over-Relaxation method to update $\phi_g$ everywhere on the
grid \cite{hirsch88}\\ For the 3D simulations, we have used the code GLOBAL
\cite{hawley95}, a 3D MHD solver that uses Eulerian finite-differencing
similar to Zeus 2D. This code was modified to include a 3D version of
the Poisson solver described above.

\subsection{Initial configuration}
Building an equilibrium self-gravitating torus is not completely
straightforward, because the density and gravitational potential influence
each other. A change in the density field modifies the gravitational
potential which in turns affects the density and so on. This suggests
the use of an iterative method. We used the self-consistent field method
developed by \inlinecite{hachisu86}.\\
Our typical model parameters
are those of a torus, with the inner and outer radii at $R_{in}=0.3$
and $R_{out}=1$ respectively. The angular velocity profile is a power
law: $\Omega \propto r^{-1.68}$ and we add a central mass $M_c$ such
that $M_c/M_d=0.5$, $M_d$ being the torus mass.  Finally, we normalized
the density such that $\rho_{max}=1$.\\
In the following, we compare the
evolution of this torus to its zero mass counterpart. The above parameters
are identical in the two cases, but the gravitational potential in the
latter is that of the central mass. This gives a density field similar
to that of the self-gravitating model, but lowers the pressure by about
an order of magnitude.

\section{2D simulations}

In this section, we describe the results of the simulations performed
in 2D with a resolution in $(R,z)=(256,256)$. A weak poloidal magnetic
field is added to the model describe above, with the toroidal component of
the vector potential being:

\begin{equation}
A_{\phi} \propto \rho \times \cos \left( 2\pi
\frac{R-R_{in}}{R_{out}-R_{in}} \right)
\label{Aphi}
\end{equation}

The components of the magnetic field are then scaled such that the volume
averaged ratio of magnetic to thermal pressure (hereafter called $\beta$)
equals $1500$ for the self-gravitating model and $200$ for the zero mass
one.\\
We found the MRI grows in both models, developing approximately the same
Maxwell stress. In the self-gravitating case, the evolution is very
similar to what was found before (see for example \opencite{hawley00}):
the early linear growth of the instability is followed by a turbulent
phase during which angular momentum is transported outward. Turbulence
then gradually decays because of the anti-dynamo theorem. During this
phase, the vertically averaged Maxwell stress tensor
$T_{R\phi}^{Max}=-\langle B_R B_{\phi}/4\pi \rho \rangle$ is dominant over the Reynolds
stress $T_{R\phi}^{Rey}=\langle v_R \delta v_{\phi}\rangle$. \\

\begin{figure}[h]
\tabcapfont
\centerline{%
\begin{tabular}{c@{\hspace{2pc}}c}
\includegraphics[scale=0.25]{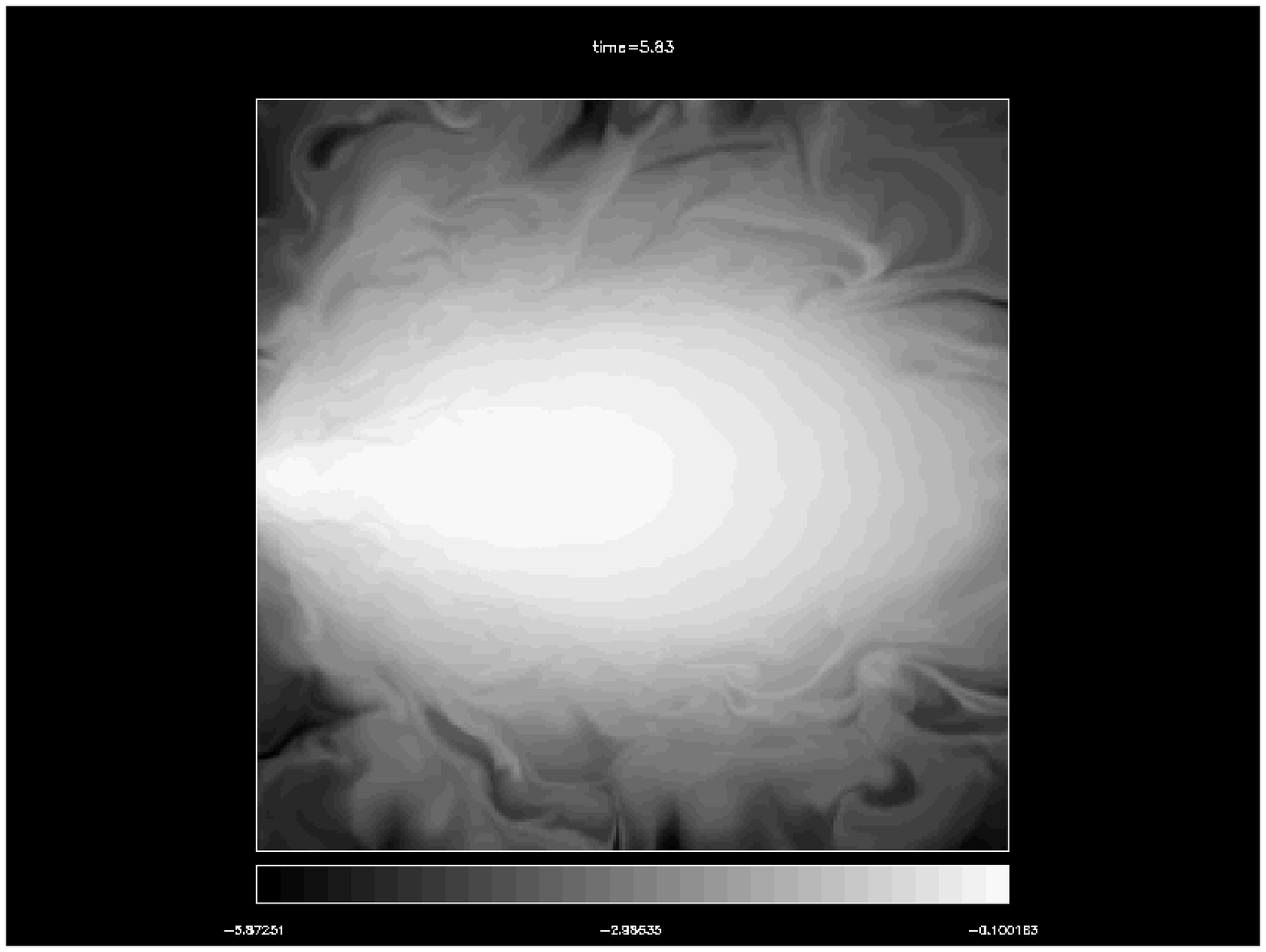} &
\includegraphics[scale=0.25]{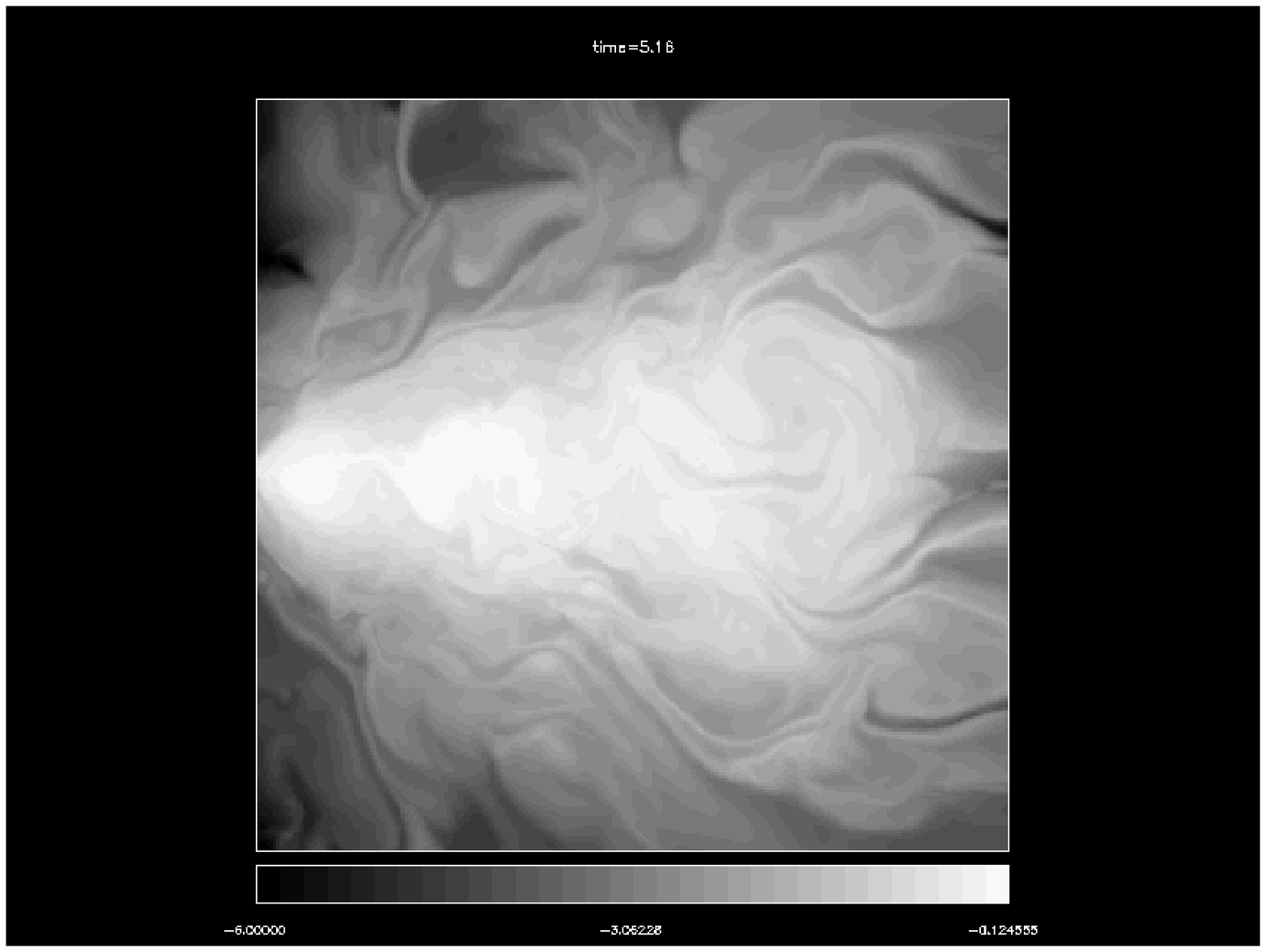} \\
a. Self-gravitating torus after 5.8 orbits & b. Zero mass torus after
5.2 orbits
\end{tabular}}
\caption{Comparison between the density logarithm distribution in the
($r$-$z$) plane for the self-gravitating torus ({\it left panel}) and the zero
mass torus ({\it right panel}). The former develops a two-component
structure composed of an inner Keplerian disks and an outer,
more massive, thick
disk. The later is disrupted by the MRI, and a standard thin disk
structure builds up, with a constant $H/R$ ratio.}
\label{snapshots}
\end{figure}

\begin{figure}
\centerline{\includegraphics[scale=0.35]{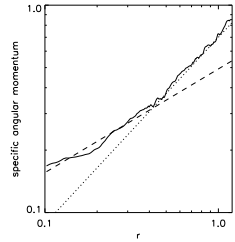}}
\caption{Angular momentum profile in the equatorial profile of the torus,
averaged between orbits 5.05 and 7. The dashed shows the Keplerian profile
resulting from the central point mass, the dotted line has a power law
dependance on the radius: $l \propto r^{0.9}$.}
\label{ang_mom}
\end{figure}

\noindent
We show in figure \ref{snapshots} the logarithm of the density field in
the $(r-z)$ plane during the turbulent phase for both models. In the
self-gravitating case ({\it left panel}), the initial torus has developed
a two-component structure, composed of an inner thin disk fed by an outer
thick and massive torus. Figure \ref{ang_mom} shows the angular momentum
radial profile in the equatorial plane during this phase ({\it solid
line}). As shown by the dashed line, the inner disk is in Keplerian
rotation around the central mass, while the dotted line is a fit of the
outer part of the disk with a power law dependence $l \propto r^{0.9}$,
very close to the Mestel profile $l \propto r$.\\   
In figure \ref{snapshots}, the comparison between both models is
striking: although the Maxwell stress is similar in the two cases, the
self-gravitating torus presents a much more coherent structure than its
zero mass counterpart. Indeed, the latter has been completely disrupted by
the initial growth of the MRI. This result is probably due to the self-gravitating potential smoothing the MRI. 

\section{3D simulations: first results}

\subsection{Hydrodynamical results}

\begin{figure}[h]
\tabcapfont
\centerline{%
\begin{tabular}{c@{\hspace{3pc}}c}
\includegraphics[scale=0.2]{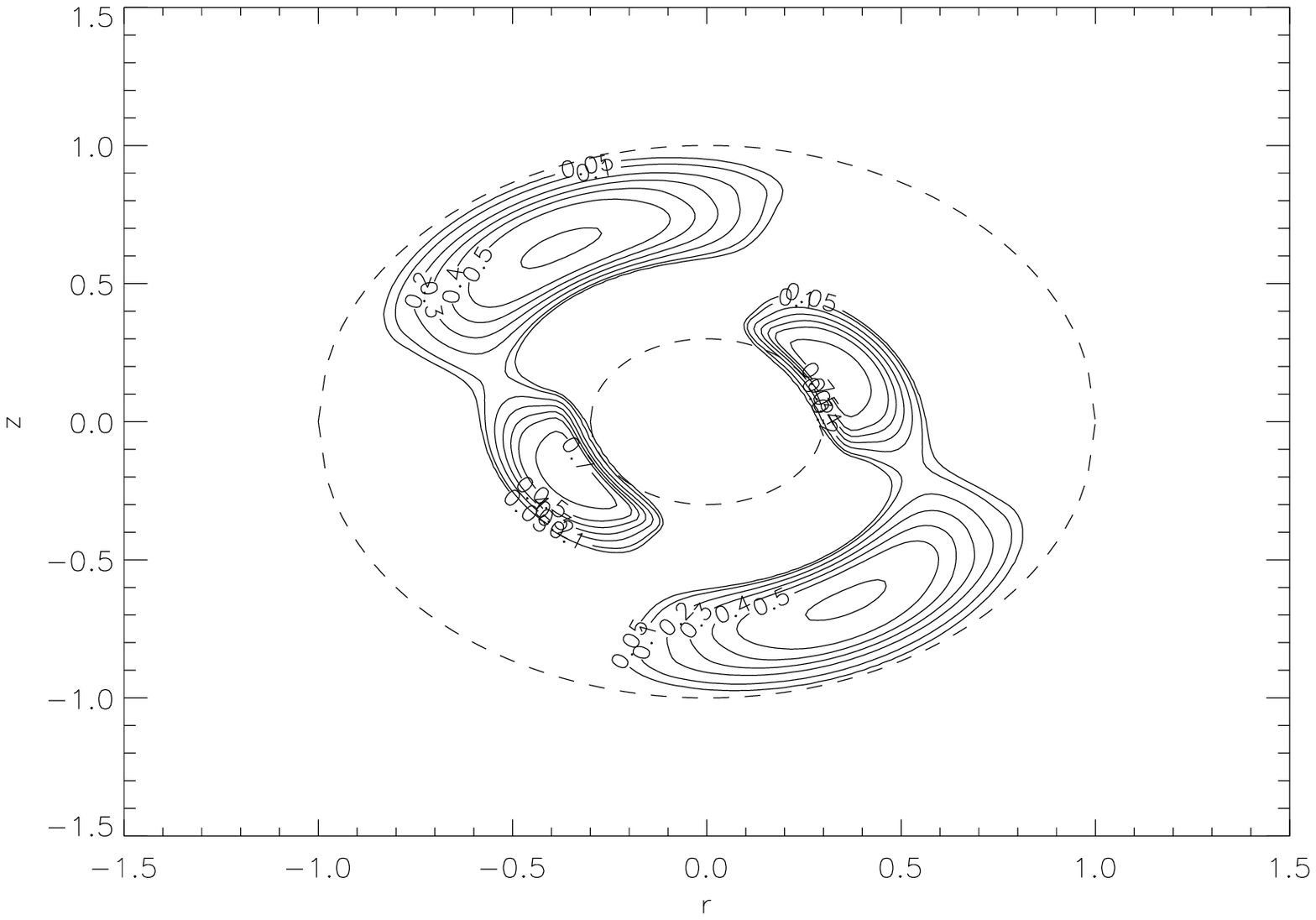} &
\includegraphics[scale=0.2]{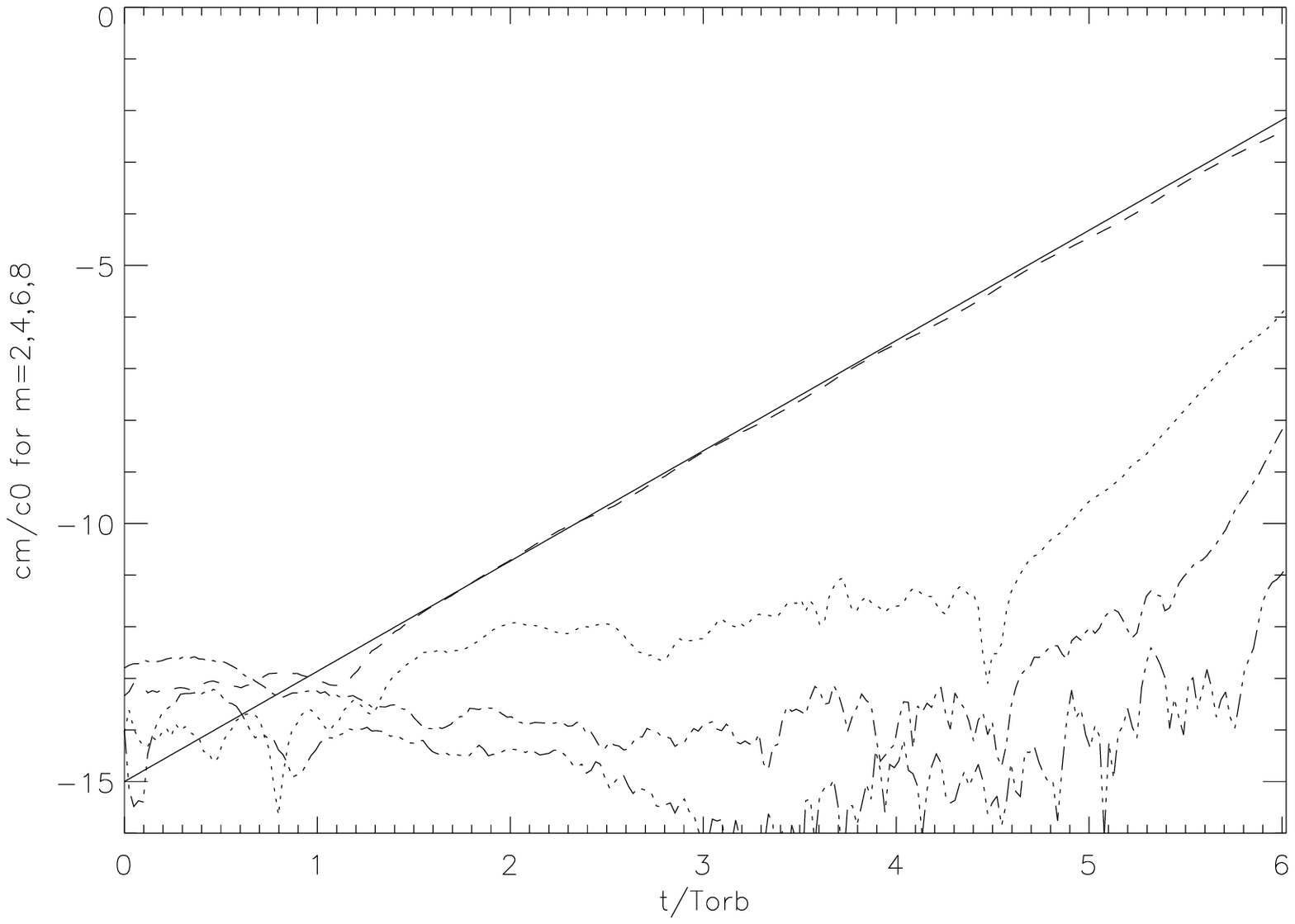}
\end{tabular}}
\caption{Hydrodynamical 3D simulations of a self-gravitating torus. The left hand side panel shows the density perturbation in the equatorial plane after 5.5 orbits. The right hand side shows the time
history of the Fourier component ({\it from top to bottom}) $m=2,4,6,8$.
An exponential growth of the $m=2$ is obvious. The measured growth rate is
similar to values previously quoted in the literature.}
\label{3D hydro}
\end{figure}

\noindent
In 3D, we expect to observe the growth of pure hydrodynamic
non-axisymmetric gravitational instabilities. To check this,
we performed nonmagnetic hydrodynamic simulations of a
torus similar to those described in the previous section, but without a
central mass. The resolution is $(N_R,N_{\theta},N_z)=(64,32,32)$ and the azimuthal range is limited to $\theta=[0,\pi]$, which prevents the appearance of odd modes. As long as the final state is not dominated by such modes, and
there is no reason to think that they will be, our qualitative results
should not be misleading.\\
The results of this run are shown in figure \ref{3D hydro}. The left panel
shows the density perturbation in the equatorial plane after $5.5$ orbits
at the initial pressure maximum. A crisp $m=2$ pattern has emerged. In the
right panel we plot the time evolution of the Fourier component
$m=2,4,6,8$. The exponential growth of the $m=2$ mode is obvious. The
growth rate measured is similar to that quoted in previous studies
\cite{tohline90}.

\subsection{MHD simulations in an axisymmetric potential}

\begin{figure}
\centerline{\includegraphics[scale=0.35]{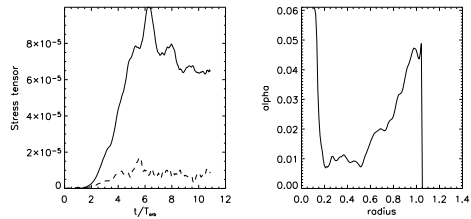}}
\caption{Left: time history of the Maxwell ({\it solid line}) and Reynolds
({\it dashed line}) stress. Right: radial profile of $\alpha$, the ratio
of the total vertically averaged stress (Maxwell + Reynolds) to the
vertically averaged pressure.}
\label{3d time history}
\end{figure}

\noindent
In 3D, different magnetic field configurations can be investigated. We
present here the results of a simulation done with an initial toroidal
field (with a the volume averaged $\beta$ of $10$). The resolution is
$(128,32,128)$ and the domain extends in $\theta$
between $0$ and $\pi/2$. However, note that this calculation is done with
only the axisymmetric component of the gravitational potential included,
preventing the growth of any non-axisymmetric {\it gravitational} instability.
Such a simplification makes the simulation much easier to do and lets us
investigate the minimum resolution required for MHD turbulence to be
sustained.\\
In the left panel of figure \ref{3d time history}, we can see that the
Maxwell stress tensor ({\it solid line}) grows similarly to 2D simulations
but saturates after about $6$ orbits and doesn't decay afterwards. We
conclude from this observation that turbulence is sustained. As in 2D, we
also see in this plot that the Reynolds stress ({\it dashed line}) is much
smaller than the Maxwell stress. The right panel shows the radial profile
of $\alpha$, the ratio of the total vertically averaged stress
(Maxwell+Reynolds) to the vertically averaged pressure. We find typical
values of the order of a few times $10^{-2}$, in agreement with previous
global non self-gravitating simulations starting with similar magnetic
configurations \cite{hawley00,steinacker02}. \\
Finally, a comparison between figure \ref{3D hydro} and \ref{3d time history} shows that both the MRI and the gravitational instabilities grow on dynamical timescales. However, it is difficult at this point to decide which of the two, if either, would dominate in the nonlinear phase, since their interaction is likely to be complex.

\section{Conclusions and Perspectives}
By means of 2D and 3D simulations, we have presented here the first
examples of the behaviour of the MRI in a global self-gravitating system.
In both 2D and 3D, we found that the MRI behaves similarly to
the zero mass
local and global configurations: the initial linear growth breaks down
into turbulence. In 2D, we observe the formation of a dual structure
composed of an inner thin disk in Keplerian rotation, fed by an outer
massive torus with a different angular momentum profile. In 3D, we
performed simulations with an axisymmetric potential and measured typical
values of $\alpha$ similar to those seen in the previous non
self-gravitating configurations.\\
A full 3D MHD simulation including high order Fourier components of
the gravitational potential is clearly needed to investigate the
interplay between the growth of non-axisymmetric instabilities (seen in
pure hydrodynamic runs) and fully developed MHD turbulence.

\end{article}

\end{document}